\documentclass[%
preprint,
superscriptaddress,
showpacs,
amsmath,amssymb,
 aps,
 prl,
]{revtex4-1}

\usepackage{graphicx}
\usepackage{dcolumn}
\usepackage{bm}

\usepackage{layouts}
\begin{document}

\preprint{}

\title{Condensation of helium in aerogels and athermal dynamics of the Random Field 
Ising Model}

\author{Geoffroy J. Aubry}%
\affiliation{%
Universit\'e Grenoble Alpes, Institut N\'eel, F-38042 Grenoble, France\\
CNRS, Institut N\'eel, F-38042 Grenoble, France
}%
\author{Fabien Bonnet}
\affiliation{%
  Universit\'e Grenoble Alpes, Institut N\'eel, F-38042 Grenoble, France\\
  CNRS, Institut N\'eel, F-38042 Grenoble, France
}%
\author{Mathieu Melich}
\affiliation{%
  Universit\'e Grenoble Alpes, Institut N\'eel, F-38042 Grenoble, France\\
  CNRS, Institut N\'eel, F-38042 Grenoble, France
}%
\author{Laurent Guyon}
\affiliation{%
  Universit\'e Grenoble Alpes, Institut N\'eel, F-38042 Grenoble, France\\
  CNRS, Institut N\'eel, F-38042 Grenoble, France
}%
\author{Panayotis Spathis }
\affiliation{%
  Universit\'e Grenoble Alpes, Institut N\'eel, F-38042 Grenoble, France\\
  CNRS, Institut N\'eel, F-38042 Grenoble, France
}%
\author{Florence Despetis }
\affiliation{
  Universit\'{e} Montpellier 2, Laboratoire Charles Coulomb UMR 5221,
  F-34095, Montpellier, France\\
  CNRS, Laboratoire Charles Coulomb UMR 5221, F-34095, Montpellier, France}
\author{Pierre-Etienne Wolf}
\email{pierre-etienne.wolf@neel.cnrs.fr}
\affiliation{%
  Universit\'e Grenoble Alpes, Institut N\'eel, F-38042 Grenoble, France\\
  CNRS, Institut N\'eel, F-38042 Grenoble, France
}%

\date{\today}

\begin{abstract}
High resolution measurements reveal that condensation isotherms of
$^4$He in a silica aerogel become discontinuous below a critical
temperature.  We show that this behaviour does not correspond to an
equilibrium phase transition modified by the disorder induced by the
aerogel structure, but to the disorder-driven critical point predicted
for the athermal out-of-equilibrium dynamics of the Random Field Ising
Model.  Our results evidence the key role of non-equilibrium effects
in the phase transitions of disordered systems.
\end{abstract}

 \pacs{ 64.60.Ht, 64.70.F-, 75.60.Ej, 67.25.bh}
\maketitle

The Ising model has a central role in statistical physics.  At the
root of the critical scaling concept in clean systems, it is also a
prototype to study the effect of disorder on phase transitions.  In
the Random Field Ising Model (RFIM), the key question is whether a
random local magnetic field acting on the Ising spins destroys the
order induced by the spin coupling.  After years of fierce theoretical
debate, it is now accepted that, in three dimensions and at low
temperature, an ordered phase should exist below a critical disorder
strength \cite{*[{For recent references, see, e.g., }] []Krzakala2010,
Fytas2013}.  The experimental situation is not that clear-cut.  In
magnetic systems, the mixing of random field and random exchange
effects makes the situation complex.  As proposed thirty years ago
\cite{Brochard1983,DeGennes1984}, the condensation (or the demixion)
of (binary) fluids in disordered porous media should offer an
alternative testing ground for the theory.  Indeed, the pure Ising
model and the liquid-vapor (or demixing) transition fall in the same
universality class, and the preferential attraction of the substrate
for one phase is equivalent to the effect of a random field.  However,
while early experiments conducted in Vycor, a porous glass, were
originally interpreted in terms of a random-field transition
\cite{Dierker1987,Wiltzius1989,Dierker1991}, the observations could
also be accounted by wetting effects in a confined, non-random,
geometry \cite{Liu1990}.  Subsequently, a breakthrough came when Wong
and Chan measured in 1990 the condensation isotherms of helium in very
porous silica aerogels, and found that their shape changed from
discontinuous to continuous at a temperature below the bulk critical
temperature \cite{Wong1990}.  The same behavior has been observed for
the condensation of nitrogen \cite{Wong1993}, and the demixion of
binary fluids inside silica gels \cite{Zhuang1996}, and similarly
interpreted in terms of the equilibrium behavior of the RFIM. But,
although these results still greatly influence present researches
\cite{Pellicane2013}, further studies demonstrated that the
condensation of helium in aerogels is reversible only very close to to
the bulk critical temperature, in which case the condensation
isotherms are always smooth \cite{Herman2005,Aubry2013}.  A steepening
of the isotherms was only observed at lower temperatures
\cite{Lambert2004a}, where condensation and evaporation are strongly
hysteretic, rising the question whether the equilibrium state is
really reached.  In fact, these observations have been suggested
\cite{Bonnet2008} to result from a totally different phenomenon,
introduced by Sethna \textit{et al.} \cite{Sethna1993,Perkovic1995}
for the RFIM, and extended to condensation of fluids in highly porous
random materials by Kierlik \textit{et al.} \cite{Kierlik2001}, namely
the occurence of an out-of-equilibrium \emph{disorder-driven critical
point}.  In this Letter, we report experiments which demonstrate the
reality of this new paradigm.

The key feature of the Sethna \textit{et al.} theory
\cite{Sethna1993,Perkovic1995} is to consider the RFIM model in the
athermal limit, where the disorder-induced barriers cannot be overcome
by thermal activation.  As the external field $H$ is swept up, the
magnetization $M$ then reverses from negative to positive by a series
of avalanches \cite{Sethna2001}, corresponding to flips of spin
clusters taking place for specific values of $H$.  Remarkably, below a
critical value for the ratio of disorder strength to spin coupling,
the largest avalanche becomes macroscopic.  Correspondingly, the
$M(H)$ curve which is smooth above the critical disorder, presents a
jump below.  Although similar to the one occurring at an \emph
{equilibrium} paramagnetic-ferromagnetic critical point, this change
of shape has a totally different origin, which Sethna \textit{et al.}
emphasize by using the term of \emph{disorder-driven critical point}.
Although in principle universal, disorder-driven critical points have
only been experimentally reported in two magnetic systems
\cite{Berger2000, Marcos2003}.  In both cases, however, the critical
exponents deduced from a scaling analysis of the $M(H)$ curves differ
from the values predicted for the 3D athermal RFIM
\cite{Perkovic1999}. As in the equilibrium case, this may result
from the presence of antiferromagnetic interactions \cite{Marcos2003}.

A Sethna \textit{et al.}-like transition has been predicted for
condensation of fluids confined in random disordered porous materials
by Kierlik \textit{et al.} \cite{Kierlik2001}.  Using a local
mean-field approach, these authors show that the analog of the
magnetisation curve, the condensation isotherm $\Phi(P)$, where $\Phi$
is the liquid fraction, and $P$ the pressure, presents a change of
shape similar to the one predicted within the athermal RFIM for
decreasing disorder~\cite{Detcheverry2003}.  It takes place when
increasing the porosity \cite{ Detcheverry2003} or decreasing the
temperature at constant, large enough,
porosity~\cite{Detcheverry2004a}.  In this paper, we observe such a
\emph{disorder-driven} transition for the condensation of helium into
a high porosity aerogel (95\% porosity, density 102~g/L) synthetized
by a one-step process at $\mathrm{pH}=9$ \cite{*[{see, e.g., }] []
Phalippou2005} \footnote{Aerogels grown in basic pH are an
experimental realization of the numerical gels studied in
ref.~\cite{Detcheverry2003, Detcheverry2004a}}.  We obtain four major
results: (i) using optical means to resolve on a local scale the
condensation process, we evidence the predicted transition from
discontinuous (vertical) to smooth isotherms; (ii) we demonstrate
unambiguously the \emph{non}-equilibrium nature of this transition;
(iii) in ageement wih theory, the occurence of the transition depends
on the pressure history; (iv) in contrast to simple expectation, the
macroscopic avalanche does not follow from some purely geometric
capillary instability \cite{DeGennes1984}.

Our aerogel was grown as a 3.7\,mm thick, 13\,mm diameter
cylinder between Teflon plates to obtain surfaces of good optical
quality, and moved to a copper cell closed by two sapphire windows.
Condensation of $^4$He is studied in an optical cryostat between
4.4\,K and the bulk critical temperature, $T_{c}\approx5.2$\,K. The
cell temperature is regulated within 10\,$\mu$K. The temperature 
heterogeneity within the aerogel resulting from the absorption of the 
300~K infrared radiation transmitted through the cryostat windows is 
estimated smaller than 1 $\mu$K \cite{Melich2011}. Blocking this IR 
radiation using shutters thermalized at 4~K has no influence on the 
isotherms, nor on our optical measurements \cite{Melich2011}. Condensation is
performed at a fixed flowrate by heating between 80\,K and 180\,K an
external reservoir connected to the cell \cite {Cross2007}.  The
typical condensation time is 10 to 20\,h, long enough for the isotherm
shape to be independent of the flowrate.  $P$ is measured at room
temperature by a pressure gauge connected to the cell by a separate
capillary, with a resolution of 10\,$\mu$bar.  The amount of $^4$He
inside the aerogel, computed by subtracting the contents of the
external reservoir and of the dead volumes from the total known amount
in the system, is converted to a global liquid fraction $\Phi$ as
described elsewhere \cite {Bonnet2008}.

\begin{figure}
  \includegraphics[width=\columnwidth]{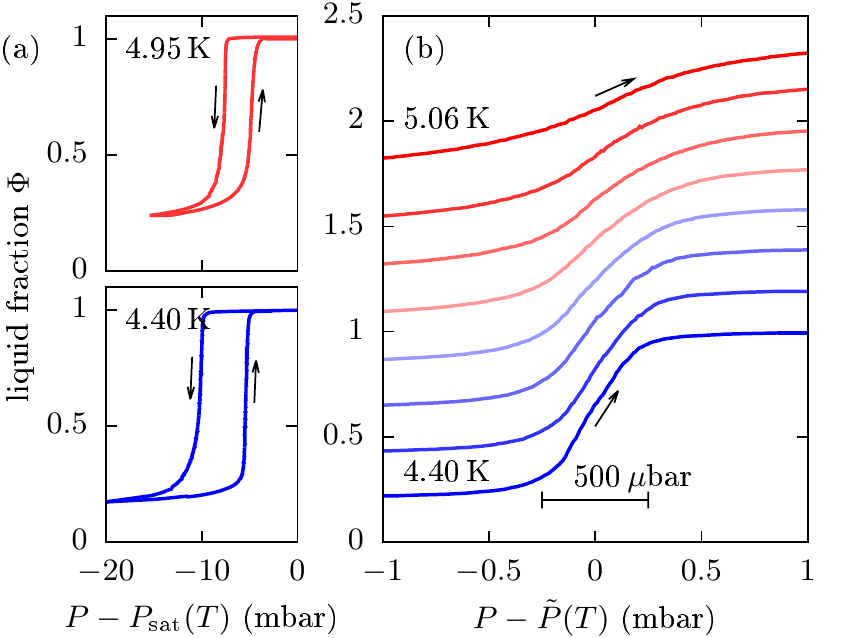}
    \caption{(color online) (a)~Hysteresis between adsorption and
    desorption isotherms at $T=4.40$\,K and 4.95\,K.
    $P_\mathrm{sat}(T)$ is the bulk vapor pressure.  (b)~Adsorption
    isotherms for $T=4.40$\,K, 4.50\,K, 4.60\,K, 4.70\,K, 4.80\,K,
    4.90\,K, 4.95\,K and 5.06\,K. $\tilde P(T)$ is the pressure at
    which the slope of the isotherm is maximum.  The $y$~scale applies
    to the 4.40\,K curve, the other curves are vertically shifted by
    0.2 for clarity.  }
  \label{fig:isotherms} 
\end{figure}

Figures~\ref{fig:isotherms}(a) and (b) show the full hysteresis loop
at 4.40\,K and 4.95\,K, and details of its condensation branch for $T$
decreasing from 5.06\,K to 4.40\,K. The isotherms become steeper as
the temperature decreases, but their slope saturates at low
temperature: at 4.40\,K, the aerogel fills over a range of
500\,$\mu$bar.  As shown by optical measurements, this finite width
results from the macroscopic heterogeneity of our sample, which masks
a disorder-driven transition occuring close to 4.7~K.
 
\begin{figure}
  \includegraphics[width=\columnwidth]{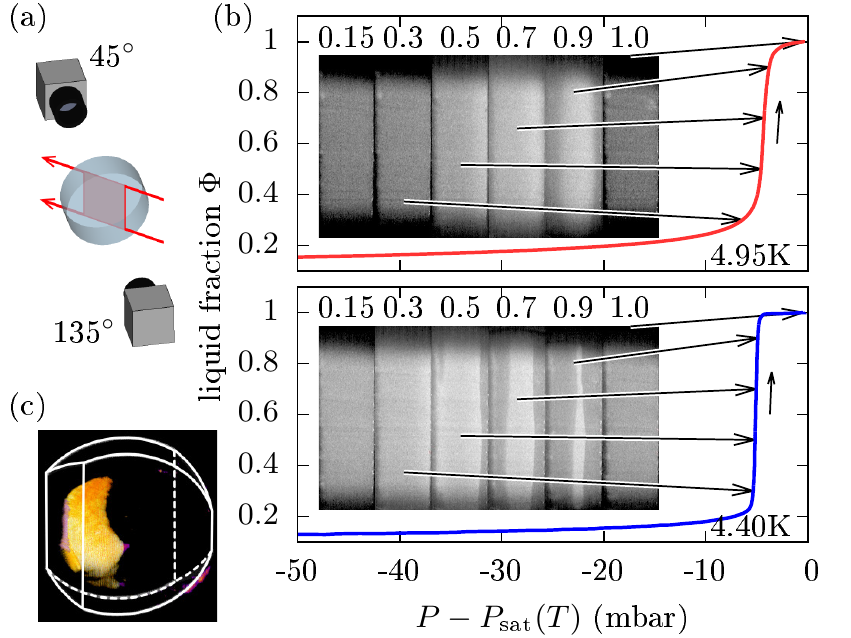}
    \caption{(color online) (a)~Positions of the laser sheet and of
    the CCD cameras relative to the aerogel.  (b)~Pictures of the
    illuminated region (1\,cm height) taken with the 135$^\circ$ CCD
    (logarithmic grayscale) at increasing liquid fractions (given
    above the pictures) during isotherms at $T=4.40$\,K and 4.95\,K.
    The arrows show the corresponding points on the isotherms.
    (c)~Perspective view of a 3D tomographic reconstruction (see text)
    close to the end of the condensation isotherm ($\Phi \approx
    0.95$) at 4.60\,K. The bright region, \textit{i.e.} the not yet
    filled part of the aerogel, is located in its left part.}
  \label{fig:thermoAndOptics} 
\end{figure}

Figure~\ref{fig:thermoAndOptics}(a) shows our optical set-up.  The
sample is illuminated with a 50\,$\mu$m thick HeNe laser sheet at a
45$^\circ$ angle with respect to the cylinder axis, and the thus
defined slice is imaged at 45$^\circ$ and 135$^\circ$ from the
incident direction.  Figure~\ref{fig:thermoAndOptics}(b) shows the
same slice for increasing values of $\Phi$ for two temperatures
\footnote{Note that the pictures only evolve when $\Phi$ increases and
not if condensation is paused.  This is similar to the spin flips of
the athermal RFIM which only take place when the field is swept.}.  In
both cases, at low liquid fraction, or for the fully filled sample,
the small and uniform brightness is due to the silica background
scattering.  When approaching the steep part of the condensation
isotherm, the scattered signal starts to grow, showing that the fluid
density becomes heterogeneous at the scale of the optical wavelength
\cite{Bonnet2008}.  At 4.40\,K, above $\Phi\approx 0.3$, the pictures
become macroscopically heterogeneous: the central bright region is
surrounded by a darker region corresponding to the fully filled state
of the aerogel.  This shows that the aerogel does not fill uniformly.
As more $^4$He is added to the cell, the bright region recesses.
Scanning the aerogel with the laser sheet evidences that the bright
region always collapses in the same outer part of the aerogel when
$\Phi\rightarrow 1$ (Fig.~\ref{fig:thermoAndOptics}(c)).  This implies
that the non uniform condensation does not result from fluid transport
from the surface to the center of the sample, but from some
heterogeneity of the sample.

\paragraph {(i) Transition from vertical to smooth isotherms\\} We
locally study the condensation process by measuring, as a function of
$P$, $I_{135}$, the signal scattered at 135$^\circ$ from a region
corresponding to one CCD pixel \footnote{The choice of 135$^\circ$
rather than 45$^\circ$ minimizes the attenuation of the scattered
light due to propagation through the bright region.}.  Taking into
account the pixel size and the laser sheet thickness, this corresponds
to a parallelepipedic region about 50\,$\mu$m wide and high, and
70\,$\mu$m deep.  This is much smaller than the sample size, but much
larger than the aerogel microscopic scales (the silica correlation
length is estimated around 10\,nm \cite{Bonnet2008}).  In practice,
because the probed region is locally homogeneous in the vertical
direction, we improve the signal-to-noise ratio by averaging $I_{135}$
in a vertical rectangle about 1\,mm high.  $I_{135}$ depends on both
the local liquid fraction $\varphi$ \emph{in the volume thus defined},
and the spatial fluid configuration.  Qualitatively, as condensation
locally proceeds, one expects liquid microdomains to appear and
increase in size, up to a point where the situation is better
described in terms of vapor microdomains decreasing in size and
number.  We thus expect $I_{135}$ to present a maximum as a function
of pressure as $\varphi$ increases from 0 to 1.

\begin{figure}
\includegraphics[width=\columnwidth]{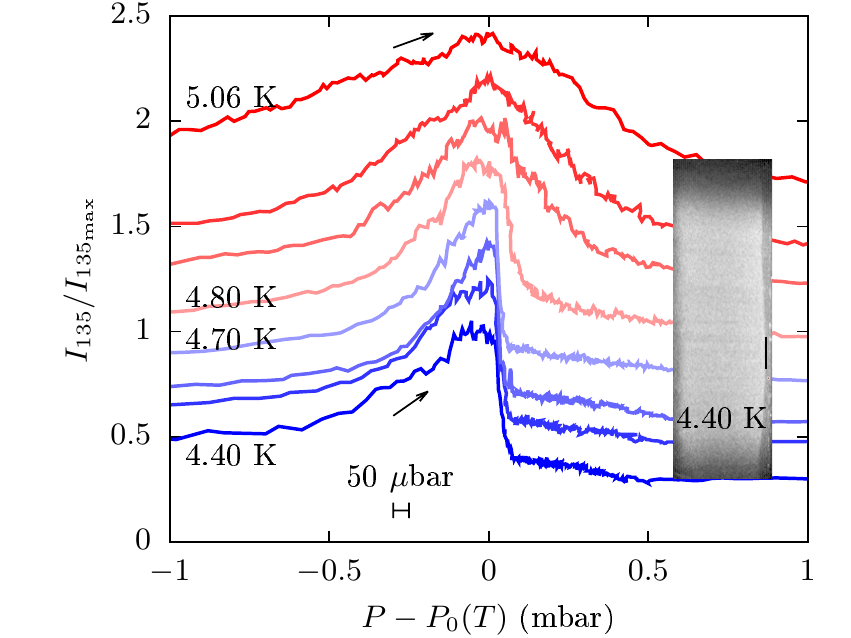}
    \caption{(color online) Intensity scattered at 135$^\circ$ (from
    the region marked by a black rectangle on the inset picture)
    during condensation at the temperatures of
    Fig.~\ref{fig:isotherms}(b).  $P_0(T)$ is the pressure at which
    the optical signal is maximum at a given temperature.  The
    $y$~scale applies to the 4.40\,K curve, the other curves are
    vertically shifted by 0.2 for clarity.}
  \label{fig:opticalSignal}
\end{figure}
The pressure dependence of $I_{135}$, shown in
Fig.~\ref{fig:opticalSignal}, evidences two different regimes of
condensation as a function of temperature.  For $T\geq 4.80$\,K,
$I_{135}$ increases as a function of pressure up to $P_0(T)$, the
pressure of its maximum, then decreases continuously, as expected for
a continuous condensation process.  In contrast, for $T\leq 4.70$\,K,
the continuous increase of $I_{135}$ up to $P_{0}$ is followed by a
sharp decrease down to the filled state value \footnote{More
precisely, the signal drops to a slightly larger value than the filled
state value.  This is a second order scattering effect, due to the
illumination of the filled aerogel by the recessing bright region.}.
The same behavior is observed at any position in the illuminated
slice, except that $P_0$ varies by about 500\,$\mu$bar across the
sample.  Convolving this spatial variation with the size of the
optically probed region accounts for the finite width
(30--40\,$\mu$bar) over which $I_{135}$ drops in
Fig.~\ref{fig:opticalSignal}.  Our observations are therefore
consistent with \emph{abrupt local} condensation isotherms below a
temperature $T^*$ ($4.70\,\mathrm{K}\leq T^*<4.80\,\mathrm{K}$).
Measurements for intermediate temperatures show that
$4.700\,\mathrm{K}\leq T^* \leq 4.725\,\mathrm{K}$ (data not shown).

\paragraph {(ii) Non-equilibrium nature of the transition\\}

\begin{figure}
  \includegraphics[page=2,width=\columnwidth]{130425_reentrance.pdf}
  \includegraphics[page=1,width=\columnwidth]{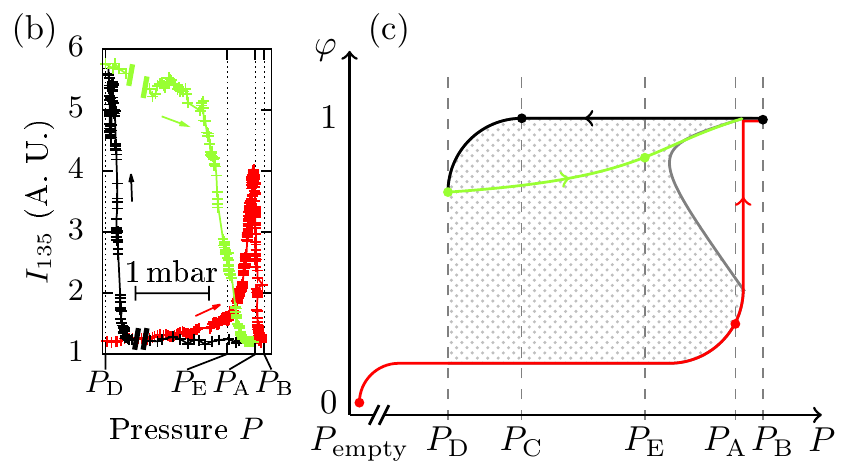}
  \caption{\label{fig:reentrance}
    (color online) (a)~Central parts of pictures taken along a minor
    hysteresis loop at $T=4.6$\,K ($<T^*$).  (b)~Corresponding optical
    signal from the dot in Fig.~\ref{fig:reentrance}(a) as a function
    of pressure.  In chronological order, dark grey/red : initial
    filling; black: partial emptying; light grey/green: filling from
    the partially emptied state.  (c)~Hypothetical distribution of
    metastable states (shaded area, only shown for $P>P_D$) in the
    $(P,\varphi)$ plane, and paths deduced from the local optical
    signal in Figs.~\ref{fig:reentrance}(a) and (b).}
\end{figure}

In the spirit of ref.~\cite{Wong1990}, this change of shape, together
with the coexistence of bright and dark regions in
Fig.~\ref{fig:thermoAndOptics}, could be interpreted in terms of a
first order \emph{equilibrium} phase transition between a gas-like
phase (inside the bright region) and the liquid phase.  In this
interpretation, $T^*$ would be the transition critical temperature,
shifted below $T_c$ due to disorder, while the observed hysteresis
would be explained by the pinning of the macroscopic interface between
the two presumed phases.  Hysteresis should then be absent
\textit{above} $T^*$, where the concept of interface has no meaning.
This is not the case as hysteresis is observed up to 5.16~K, well
above $T^*$~\cite{Aubry2013}.  Moreover, optical measurements show that the fluid
distribution in the gas-like region is not the same on condensation
and evaporation, directly showing that this region does not correspond
to a true thermodynamic phase.  This is illustrated by
figure~\ref{fig:reentrance}(a) which shows pictures taken at 4.60\,K
($<T^*$) when reverting from condensation to evaporation before the
aerogel is fully filled.  Starting from the empty state, we condense
up to the pressure $P_\mathrm{B}$, where the dot in
Fig.~\ref{fig:reentrance}(a) is just filled (picture B), then
evaporate by decreasing the pressure.  In the interpretation of a
pinned interface, one would expect, once the pressure reaches the
pressure $P_\mathrm{D}$ of the evaporation branch of the full
hysteresis loop, the interface to move backwards, retracing in
reverted order the same states than during filling.  This is not
observed.  At $P_\mathrm{D}$, when the dot starts to empty, its state
differs from the state upon filling, as indicated by the increased
local brightness (picture~D and fig.~\ref{fig:reentrance}(b)).  This
set of results excludes that the abrupt condensation below $T^{*}$
corresponds to an equilibrium phase transition.

\paragraph {(iii) History dependence\\} We now identify this
phenomenon to the out-of-equilibrium transition predicted by Sethna
\textit{et al.} by studying its dependence on history.  From results
on the athermal RFIM on a random graph \cite{Perez-Reche2008}, one
expects that, starting from an empty aerogel, a macroscopic jump of
$\varphi$ follows from a reentrance of the outer envelope of the
distribution of metastable states in the $(P,\varphi)$ plane, as, for
this initial condition, condensation proceeds along this envelope.
For such a distribution, illustrated in fig.~\ref{fig:reentrance}(c),
filling from an only partly emptied initial state should suppress the
jump.  As shown by figure~\ref{fig:reentrance}(b), this is indeed the
behavior observed for the dot of figure~\ref{fig:reentrance}(a): When
starting from $P_\mathrm{D}$, where it was only partly emptied, the
dot continuously fills (its scattered signal continuously decreases to
the liquid state value, unlike what observed when filling from the
fully emptied state.)

\paragraph {(iv) Microscopic origin of the abrupt condensation\\}

At 4.7~K, liquid and gas can be discriminated on the scale of the
aerogel correlation length.  One could then interpret the abrupt
condensation below $T^{*}$ as a capillary instability of the
microscopic liquid-vapor meniscii connecting neighboring silica
strands, a scenario first introduced by de Gennes \cite{DeGennes1984}.
The abrupt condensation event should then take place for a given
meniscii configuration, independently on temperature, hence for a
given average size of the liquid microdomains.  This is not the case.
While, at a given temperature, the value of $I_{135}$ just before full
condensation (\textit{i.e.} the value of $I_{135_\mathrm{max}}$ in
Fig.~\ref {fig:opticalSignal}) is the same across the aerogel width,
consistent with the macroscopic avalanche occurring for a given
average size of the liquid microdomains, this size depends on
temperature.  To show it, we have measured the scattering anisotropy,
as given by the ratio of the intensities scattered at 45$^\circ$ and
135$^\circ$, after subtraction of the silica background \footnote{This
quantity is measured, without any error due to attenuation, at the
last pressure where the bright domain intersects the illuminated
section.}.  This ratio increases from 2 at 4.40\,K up to 5 at 4.70\,K,
implying that the size of the liquid microdomains at the avalanche
threshold increases with temperature.  This behavior is in qualitative
agreement with preliminary simulations within the Kierlik \textit{et
al.}'s model~\cite{Aubry2013b}.

\paragraph {Conclusion\\}

Our measurements demonstrate that, despite the finite temperature, the
concept of athermal, out-of-equilibrium, dynamics accounts for the
process of fluid condensation into our 95\% porosity aerogel.  We
confirmed the universal character of this concept by measurements on a
lighter (98.5\%) aerogel~\footnote{See Supplemental Material at [URL] for 
a description of the results on this light aerogel.}.
The larger value of $T^*$ measured in this case agrees with the
theoretical expectation that $T^*$ increases when disorder
decreases~\cite{Detcheverry2005a}.  Our results lead to revisit
earlier experiments \cite{Wong1990,Wong1993,Zhuang1996} where a change
of the isotherms shape has been interpreted in terms of a
disorder-shifted equilibrium critical point.  Similarly to our case,
it could rather result from an out-of-equilibrium critical point.  As
hysteresis occurs on both sides of $T^*$ in the out-of-equilibrium
scenario, this would naturally explain the slow dynamics reported
above $T^*$ in these experiments.  While our results thus suggest that
observing a true equilibrium behavior in RFIM-like disordered systems
could be challenging, they also show the pertinence of non-equilibrium
concepts to describe the properties of first-order phase transitions
in presence of disorder.

We acknowledge fruitful
discussions with J.C.~Angl\`es~d'Auriac, E.~Kierlik, and
M.~L.~Rosinberg, and support from ANR-06-BLAN-0098.

\bibliography{}

\begin{thebibliography}{37}%
\makeatletter
\providecommand \@ifxundefined [1]{%
 \@ifx{#1\undefined}
}%
\providecommand \@ifnum [1]{%
 \ifnum #1\expandafter \@firstoftwo
 \else \expandafter \@secondoftwo
 \fi
}%
\providecommand \@ifx [1]{%
 \ifx #1\expandafter \@firstoftwo
 \else \expandafter \@secondoftwo
 \fi
}%
\providecommand \natexlab [1]{#1}%
\providecommand \enquote  [1]{``#1''}%
\providecommand \bibnamefont  [1]{#1}%
\providecommand \bibfnamefont [1]{#1}%
\providecommand \citenamefont [1]{#1}%
\providecommand \href@noop [0]{\@secondoftwo}%
\providecommand \href [0]{\begingroup \@sanitize@url \@href}%
\providecommand \@href[1]{\@@startlink{#1}\@@href}%
\providecommand \@@href[1]{\endgroup#1\@@endlink}%
\providecommand \@sanitize@url [0]{\catcode `\\12\catcode `\$12\catcode
  `\&12\catcode `\#12\catcode `\^12\catcode `\_12\catcode `\%12\relax}%
\providecommand \@@startlink[1]{}%
\providecommand \@@endlink[0]{}%
\providecommand \url  [0]{\begingroup\@sanitize@url \@url }%
\providecommand \@url [1]{\endgroup\@href {#1}{\urlprefix }}%
\providecommand \urlprefix  [0]{URL }%
\providecommand \Eprint [0]{\href }%
\providecommand \doibase [0]{http://dx.doi.org/}%
\providecommand \selectlanguage [0]{\@gobble}%
\providecommand \bibinfo  [0]{\@secondoftwo}%
\providecommand \bibfield  [0]{\@secondoftwo}%
\providecommand \translation [1]{[#1]}%
\providecommand \BibitemOpen [0]{}%
\providecommand \bibitemStop [0]{}%
\providecommand \bibitemNoStop [0]{.\EOS\space}%
\providecommand \EOS [0]{\spacefactor3000\relax}%
\providecommand \BibitemShut  [1]{\csname bibitem#1\endcsname}%
\let\auto@bib@innerbib\@empty
\bibitem [{\citenamefont {Krzakala}\ \emph {et~al.}(2010)\citenamefont
  {Krzakala}, \citenamefont {Ricci-Tersenghi},\ and\ \citenamefont
  {Zdeborov\'a}}]{Krzakala2010}%
  \BibitemOpen
  \bibfield  {author} {\bibinfo {author} {\bibfnamefont {F.}~\bibnamefont
  {Krzakala}}, \bibinfo {author} {\bibfnamefont {F.}~\bibnamefont
  {Ricci-Tersenghi}}, \ and\ \bibinfo {author} {\bibfnamefont {L.}~\bibnamefont
  {Zdeborov\'a}},\ }\href {\doibase 10.1103/PhysRevLett.104.207208} {\bibfield
  {journal} {\bibinfo  {journal} {Phys. Rev. Lett.}\ }\textbf {\bibinfo
  {volume} {104}},\ \bibinfo {pages} {207208} (\bibinfo {year}
  {2010})}\BibitemShut {NoStop}%
\bibitem [{\citenamefont {Fytas}\ and\ \citenamefont
  {Mart\'in-Mayor}(2013)}]{Fytas2013}%
  \BibitemOpen
  \bibfield  {author} {\bibinfo {author} {\bibfnamefont {N.~G.}\ \bibnamefont
  {Fytas}}\ and\ \bibinfo {author} {\bibfnamefont {V.}~\bibnamefont
  {Mart\'in-Mayor}},\ }\href {\doibase 10.1103/PhysRevLett.110.227201}
  {\bibfield  {journal} {\bibinfo  {journal} {Phys. Rev. Lett.}\ }\textbf
  {\bibinfo {volume} {110}},\ \bibinfo {pages} {227201} (\bibinfo {year}
  {2013})}\BibitemShut {NoStop}%
\bibitem [{\citenamefont {Brochard}\ and\ \citenamefont
  {de~Gennes}(1983)}]{Brochard1983}%
  \BibitemOpen
  \bibfield  {author} {\bibinfo {author} {\bibfnamefont {F.}~\bibnamefont
  {Brochard}}\ and\ \bibinfo {author} {\bibfnamefont {P.~G.}\ \bibnamefont
  {de~Gennes}},\ }\href {\doibase 10.1051/jphyslet:019830044018078500}
  {\bibfield  {journal} {\bibinfo  {journal} {J. Physique Lett.}\ }\textbf
  {\bibinfo {volume} {44}},\ \bibinfo {pages} {785} (\bibinfo {year}
  {1983})}\BibitemShut {NoStop}%
\bibitem [{\citenamefont {de~Gennes}(1984)}]{DeGennes1984}%
  \BibitemOpen
  \bibfield  {author} {\bibinfo {author} {\bibfnamefont {P.~G.}\ \bibnamefont
  {de~Gennes}},\ }\href {\doibase 10.1021/j150670a004} {\bibfield  {journal}
  {\bibinfo  {journal} {The Journal of Physical Chemistry}\ }\textbf {\bibinfo
  {volume} {88}},\ \bibinfo {pages} {6469} (\bibinfo {year}
  {1984})}\BibitemShut {NoStop}%
\bibitem [{\citenamefont {Dierker}\ and\ \citenamefont
  {Wiltzius}(1987)}]{Dierker1987}%
  \BibitemOpen
  \bibfield  {author} {\bibinfo {author} {\bibfnamefont {S.~B.}\ \bibnamefont
  {Dierker}}\ and\ \bibinfo {author} {\bibfnamefont {P.}~\bibnamefont
  {Wiltzius}},\ }\href {\doibase 10.1103/PhysRevLett.58.1865} {\bibfield
  {journal} {\bibinfo  {journal} {Phys. Rev. Lett.}\ }\textbf {\bibinfo
  {volume} {58}},\ \bibinfo {pages} {1865} (\bibinfo {year}
  {1987})}\BibitemShut {NoStop}%
\bibitem [{\citenamefont {Wiltzius}\ \emph {et~al.}(1989)\citenamefont
  {Wiltzius}, \citenamefont {Dierker},\ and\ \citenamefont
  {Dennis}}]{Wiltzius1989}%
  \BibitemOpen
  \bibfield  {author} {\bibinfo {author} {\bibfnamefont {P.}~\bibnamefont
  {Wiltzius}}, \bibinfo {author} {\bibfnamefont {S.~B.}\ \bibnamefont
  {Dierker}}, \ and\ \bibinfo {author} {\bibfnamefont {B.~S.}\ \bibnamefont
  {Dennis}},\ }\href {\doibase 10.1103/PhysRevLett.62.804} {\bibfield
  {journal} {\bibinfo  {journal} {Phys. Rev. Lett.}\ }\textbf {\bibinfo
  {volume} {62}},\ \bibinfo {pages} {804} (\bibinfo {year} {1989})}\BibitemShut
  {NoStop}%
\bibitem [{\citenamefont {Dierker}\ and\ \citenamefont
  {Wiltzius}(1991)}]{Dierker1991}%
  \BibitemOpen
  \bibfield  {author} {\bibinfo {author} {\bibfnamefont {S.~B.}\ \bibnamefont
  {Dierker}}\ and\ \bibinfo {author} {\bibfnamefont {P.}~\bibnamefont
  {Wiltzius}},\ }\href {\doibase 10.1103/PhysRevLett.66.1185} {\bibfield
  {journal} {\bibinfo  {journal} {Phys. Rev. Lett.}\ }\textbf {\bibinfo
  {volume} {66}},\ \bibinfo {pages} {1185} (\bibinfo {year}
  {1991})}\BibitemShut {NoStop}%
\bibitem [{\citenamefont {Liu}\ \emph {et~al.}(1990)\citenamefont {Liu},
  \citenamefont {Durian}, \citenamefont {Herbolzheimer},\ and\ \citenamefont
  {Safran}}]{Liu1990}%
  \BibitemOpen
  \bibfield  {author} {\bibinfo {author} {\bibfnamefont {A.~J.}\ \bibnamefont
  {Liu}}, \bibinfo {author} {\bibfnamefont {D.~J.}\ \bibnamefont {Durian}},
  \bibinfo {author} {\bibfnamefont {E.}~\bibnamefont {Herbolzheimer}}, \ and\
  \bibinfo {author} {\bibfnamefont {S.~A.}\ \bibnamefont {Safran}},\ }\href
  {\doibase 10.1103/PhysRevLett.65.1897} {\bibfield  {journal} {\bibinfo
  {journal} {Phys. Rev. Lett.}\ }\textbf {\bibinfo {volume} {65}},\ \bibinfo
  {pages} {1897} (\bibinfo {year} {1990})}\BibitemShut {NoStop}%
\bibitem [{\citenamefont {Wong}\ and\ \citenamefont {Chan}(1990)}]{Wong1990}%
  \BibitemOpen
  \bibfield  {author} {\bibinfo {author} {\bibfnamefont {A.~P.~Y.}\
  \bibnamefont {Wong}}\ and\ \bibinfo {author} {\bibfnamefont {M.~H.~W.}\
  \bibnamefont {Chan}},\ }\href {\doibase 10.1103/PhysRevLett.65.2567}
  {\bibfield  {journal} {\bibinfo  {journal} {Phys. Rev. Lett.}\ }\textbf
  {\bibinfo {volume} {65}},\ \bibinfo {pages} {2567} (\bibinfo {year}
  {1990})}\BibitemShut {NoStop}%
\bibitem [{\citenamefont {Wong}\ \emph {et~al.}(1993)\citenamefont {Wong},
  \citenamefont {Kim}, \citenamefont {Goldburg},\ and\ \citenamefont
  {Chan}}]{Wong1993}%
  \BibitemOpen
  \bibfield  {author} {\bibinfo {author} {\bibfnamefont {A.~P.~Y.}\
  \bibnamefont {Wong}}, \bibinfo {author} {\bibfnamefont {S.~B.}\ \bibnamefont
  {Kim}}, \bibinfo {author} {\bibfnamefont {W.~I.}\ \bibnamefont {Goldburg}}, \
  and\ \bibinfo {author} {\bibfnamefont {M.~H.~W.}\ \bibnamefont {Chan}},\
  }\href {\doibase 10.1103/PhysRevLett.70.954} {\bibfield  {journal} {\bibinfo
  {journal} {Phys. Rev. Lett.}\ }\textbf {\bibinfo {volume} {70}},\ \bibinfo
  {pages} {954} (\bibinfo {year} {1993})}\BibitemShut {NoStop}%
\bibitem [{\citenamefont {Zhuang}\ \emph {et~al.}(1996)\citenamefont {Zhuang},
  \citenamefont {Casielles},\ and\ \citenamefont {Cannell}}]{Zhuang1996}%
  \BibitemOpen
  \bibfield  {author} {\bibinfo {author} {\bibfnamefont {Z.}~\bibnamefont
  {Zhuang}}, \bibinfo {author} {\bibfnamefont {A.~G.}\ \bibnamefont
  {Casielles}}, \ and\ \bibinfo {author} {\bibfnamefont {D.~S.}\ \bibnamefont
  {Cannell}},\ }\href {\doibase 10.1103/PhysRevLett.77.2969} {\bibfield
  {journal} {\bibinfo  {journal} {Phys. Rev. Lett.}\ }\textbf {\bibinfo
  {volume} {77}},\ \bibinfo {pages} {2969} (\bibinfo {year}
  {1996})}\BibitemShut {NoStop}%
\bibitem [{\citenamefont {Pellicane}\ \emph {et~al.}(2013)\citenamefont
  {Pellicane}, \citenamefont {Vink}, \citenamefont {Russo},\ and\ \citenamefont
  {Giaquinta}}]{Pellicane2013}%
  \BibitemOpen
  \bibfield  {author} {\bibinfo {author} {\bibfnamefont {G.}~\bibnamefont
  {Pellicane}}, \bibinfo {author} {\bibfnamefont {R.~L.~C.}\ \bibnamefont
  {Vink}}, \bibinfo {author} {\bibfnamefont {B.}~\bibnamefont {Russo}}, \ and\
  \bibinfo {author} {\bibfnamefont {P.~V.}\ \bibnamefont {Giaquinta}},\ }\href
  {\doibase 10.1103/PhysRevE.88.042131} {\bibfield  {journal} {\bibinfo
  {journal} {Phys. Rev. E}\ }\textbf {\bibinfo {volume} {88}},\ \bibinfo
  {pages} {042131} (\bibinfo {year} {2013})}\BibitemShut {NoStop}%
\bibitem [{\citenamefont {Herman}\ \emph {et~al.}(2005)\citenamefont {Herman},
  \citenamefont {Day},\ and\ \citenamefont {Beamish}}]{Herman2005}%
  \BibitemOpen
  \bibfield  {author} {\bibinfo {author} {\bibfnamefont {T.}~\bibnamefont
  {Herman}}, \bibinfo {author} {\bibfnamefont {J.}~\bibnamefont {Day}}, \ and\
  \bibinfo {author} {\bibfnamefont {J.}~\bibnamefont {Beamish}},\ }\href
  {\doibase 10.1103/PhysRevB.72.184202} {\bibfield  {journal} {\bibinfo
  {journal} {Phys. Rev. B}\ }\textbf {\bibinfo {volume} {72}},\ \bibinfo
  {pages} {184202} (\bibinfo {year} {2005})}\BibitemShut {NoStop}%
\bibitem [{\citenamefont {Aubry}\ \emph {et~al.}(2013)\citenamefont {Aubry},
  \citenamefont {Bonnet}, \citenamefont {Melich}, \citenamefont {Guyon},
  \citenamefont {Despetis},\ and\ \citenamefont {Wolf}}]{Aubry2013}%
  \BibitemOpen
  \bibfield  {author} {\bibinfo {author} {\bibfnamefont {G.~J.}\ \bibnamefont
  {Aubry}}, \bibinfo {author} {\bibfnamefont {F.}~\bibnamefont {Bonnet}},
  \bibinfo {author} {\bibfnamefont {M.}~\bibnamefont {Melich}}, \bibinfo
  {author} {\bibfnamefont {L.}~\bibnamefont {Guyon}}, \bibinfo {author}
  {\bibfnamefont {F.}~\bibnamefont {Despetis}}, \ and\ \bibinfo {author}
  {\bibfnamefont {P.~E.}\ \bibnamefont {Wolf}},\ }\href {\doibase
  10.1007/s10909-012-0740-2} {\bibfield  {journal} {\bibinfo  {journal} {J. Low
  Temp. Phys.}\ }\textbf {\bibinfo {volume} {171}},\ \bibinfo {pages} {670}
  (\bibinfo {year} {2013})}\BibitemShut {NoStop}%
\bibitem [{\citenamefont {Lambert}\ \emph {et~al.}(2004)\citenamefont
  {Lambert}, \citenamefont {Gabay}, \citenamefont {Puech},\ and\ \citenamefont
  {Wolf}}]{Lambert2004a}%
  \BibitemOpen
  \bibfield  {author} {\bibinfo {author} {\bibfnamefont {T.}~\bibnamefont
  {Lambert}}, \bibinfo {author} {\bibfnamefont {C.}~\bibnamefont {Gabay}},
  \bibinfo {author} {\bibfnamefont {L.}~\bibnamefont {Puech}}, \ and\ \bibinfo
  {author} {\bibfnamefont {P.~E.}\ \bibnamefont {Wolf}},\ }\href {\doibase
  10.1023/B:JOLT.0000012569.91725.5d} {\bibfield  {journal} {\bibinfo
  {journal} {J. Low Temp. Phys.}\ }\textbf {\bibinfo {volume} {134}},\ \bibinfo
  {pages} {293} (\bibinfo {year} {2004})}\BibitemShut {NoStop}%
\bibitem [{\citenamefont {Bonnet}\ \emph {et~al.}(2008)\citenamefont {Bonnet},
  \citenamefont {Lambert}, \citenamefont {Cross}, \citenamefont {Guyon},
  \citenamefont {Despetis}, \citenamefont {Puech},\ and\ \citenamefont
  {Wolf}}]{Bonnet2008}%
  \BibitemOpen
  \bibfield  {author} {\bibinfo {author} {\bibfnamefont {F.}~\bibnamefont
  {Bonnet}}, \bibinfo {author} {\bibfnamefont {T.}~\bibnamefont {Lambert}},
  \bibinfo {author} {\bibfnamefont {B.}~\bibnamefont {Cross}}, \bibinfo
  {author} {\bibfnamefont {L.}~\bibnamefont {Guyon}}, \bibinfo {author}
  {\bibfnamefont {F.}~\bibnamefont {Despetis}}, \bibinfo {author}
  {\bibfnamefont {L.}~\bibnamefont {Puech}}, \ and\ \bibinfo {author}
  {\bibfnamefont {P.~E.}\ \bibnamefont {Wolf}},\ }\href {\doibase
  10.1209/0295-5075/82/56003} {\bibfield  {journal} {\bibinfo  {journal}
  {Europhys. Lett.}\ }\textbf {\bibinfo {volume} {82}},\ \bibinfo {pages}
  {56003} (\bibinfo {year} {2008})},\ \Eprint {http://arxiv.org/abs/0802.2985}
  {arXiv:0802.2985 [cond-mat.soft]} \BibitemShut {NoStop}%
\bibitem [{\citenamefont {Sethna}\ \emph {et~al.}(1993)\citenamefont {Sethna},
  \citenamefont {Dahmen}, \citenamefont {Kartha}, \citenamefont {Krumhansl},
  \citenamefont {Roberts},\ and\ \citenamefont {Shore}}]{Sethna1993}%
  \BibitemOpen
  \bibfield  {author} {\bibinfo {author} {\bibfnamefont {J.~P.}\ \bibnamefont
  {Sethna}}, \bibinfo {author} {\bibfnamefont {K.}~\bibnamefont {Dahmen}},
  \bibinfo {author} {\bibfnamefont {S.}~\bibnamefont {Kartha}}, \bibinfo
  {author} {\bibfnamefont {J.~A.}\ \bibnamefont {Krumhansl}}, \bibinfo {author}
  {\bibfnamefont {B.~W.}\ \bibnamefont {Roberts}}, \ and\ \bibinfo {author}
  {\bibfnamefont {J.~D.}\ \bibnamefont {Shore}},\ }\href {\doibase
  10.1103/PhysRevLett.70.3347} {\bibfield  {journal} {\bibinfo  {journal}
  {Phys. Rev. Lett.}\ }\textbf {\bibinfo {volume} {70}},\ \bibinfo {pages}
  {3347} (\bibinfo {year} {1993})}\BibitemShut {NoStop}%
\bibitem [{\citenamefont {Perkovi\ifmmode~\acute{c}\else \'{c}\fi{}}\ \emph
  {et~al.}(1995)\citenamefont {Perkovi\ifmmode~\acute{c}\else \'{c}\fi{}},
  \citenamefont {Dahmen},\ and\ \citenamefont {Sethna}}]{Perkovic1995}%
  \BibitemOpen
  \bibfield  {author} {\bibinfo {author} {\bibfnamefont {O.}~\bibnamefont
  {Perkovi\ifmmode~\acute{c}\else \'{c}\fi{}}}, \bibinfo {author}
  {\bibfnamefont {K.}~\bibnamefont {Dahmen}}, \ and\ \bibinfo {author}
  {\bibfnamefont {J.~P.}\ \bibnamefont {Sethna}},\ }\href {\doibase
  10.1103/PhysRevLett.75.4528} {\bibfield  {journal} {\bibinfo  {journal}
  {Phys. Rev. Lett.}\ }\textbf {\bibinfo {volume} {75}},\ \bibinfo {pages}
  {4528} (\bibinfo {year} {1995})}\BibitemShut {NoStop}%
\bibitem [{\citenamefont {Kierlik}\ \emph {et~al.}(2001)\citenamefont
  {Kierlik}, \citenamefont {Monson}, \citenamefont {Rosinberg}, \citenamefont
  {Sarkisov},\ and\ \citenamefont {Tarjus}}]{Kierlik2001}%
  \BibitemOpen
  \bibfield  {author} {\bibinfo {author} {\bibfnamefont {E.}~\bibnamefont
  {Kierlik}}, \bibinfo {author} {\bibfnamefont {P.~A.}\ \bibnamefont {Monson}},
  \bibinfo {author} {\bibfnamefont {M.~L.}\ \bibnamefont {Rosinberg}}, \bibinfo
  {author} {\bibfnamefont {L.}~\bibnamefont {Sarkisov}}, \ and\ \bibinfo
  {author} {\bibfnamefont {G.}~\bibnamefont {Tarjus}},\ }\href {\doibase
  10.1103/PhysRevLett.87.055701} {\bibfield  {journal} {\bibinfo  {journal}
  {Phys. Rev. Lett.}\ }\textbf {\bibinfo {volume} {87}},\ \bibinfo {pages}
  {055701} (\bibinfo {year} {2001})}\BibitemShut {NoStop}%
\bibitem [{\citenamefont {Sethna}\ \emph {et~al.}(2001)\citenamefont {Sethna},
  \citenamefont {Dahmen},\ and\ \citenamefont {Myers}}]{Sethna2001}%
  \BibitemOpen
  \bibfield  {author} {\bibinfo {author} {\bibfnamefont {J.~P.}\ \bibnamefont
  {Sethna}}, \bibinfo {author} {\bibfnamefont {K.~A.}\ \bibnamefont {Dahmen}},
  \ and\ \bibinfo {author} {\bibfnamefont {C.~R.}\ \bibnamefont {Myers}},\
  }\href {\doibase 10.1038/35065675} {\bibfield  {journal} {\bibinfo  {journal}
  {Nature (London)}\ }\textbf {\bibinfo {volume} {410}},\ \bibinfo {pages}
  {242} (\bibinfo {year} {2001})}\BibitemShut {NoStop}%
\bibitem [{\citenamefont {Berger}\ \emph {et~al.}(2000)\citenamefont {Berger},
  \citenamefont {Inomata}, \citenamefont {Jiang}, \citenamefont {Pearson},\
  and\ \citenamefont {Bader}}]{Berger2000}%
  \BibitemOpen
  \bibfield  {author} {\bibinfo {author} {\bibfnamefont {A.}~\bibnamefont
  {Berger}}, \bibinfo {author} {\bibfnamefont {A.}~\bibnamefont {Inomata}},
  \bibinfo {author} {\bibfnamefont {J.~S.}\ \bibnamefont {Jiang}}, \bibinfo
  {author} {\bibfnamefont {J.~E.}\ \bibnamefont {Pearson}}, \ and\ \bibinfo
  {author} {\bibfnamefont {S.~D.}\ \bibnamefont {Bader}},\ }\href {\doibase
  10.1103/PhysRevLett.85.4176} {\bibfield  {journal} {\bibinfo  {journal}
  {Phys. Rev. Lett.}\ }\textbf {\bibinfo {volume} {85}},\ \bibinfo {pages}
  {4176} (\bibinfo {year} {2000})}\BibitemShut {NoStop}%
\bibitem [{\citenamefont {Marcos}\ \emph {et~al.}(2003)\citenamefont {Marcos},
  \citenamefont {Vives}, \citenamefont {Ma\~nosa}, \citenamefont {Acet},
  \citenamefont {Duman}, \citenamefont {Morin}, \citenamefont {Nov\'ak},\ and\
  \citenamefont {Planes}}]{Marcos2003}%
  \BibitemOpen
  \bibfield  {author} {\bibinfo {author} {\bibfnamefont {J.}~\bibnamefont
  {Marcos}}, \bibinfo {author} {\bibfnamefont {E.}~\bibnamefont {Vives}},
  \bibinfo {author} {\bibfnamefont {L.}~\bibnamefont {Ma\~nosa}}, \bibinfo
  {author} {\bibfnamefont {M.}~\bibnamefont {Acet}}, \bibinfo {author}
  {\bibfnamefont {E.}~\bibnamefont {Duman}}, \bibinfo {author} {\bibfnamefont
  {M.}~\bibnamefont {Morin}}, \bibinfo {author} {\bibfnamefont
  {V.}~\bibnamefont {Nov\'ak}}, \ and\ \bibinfo {author} {\bibfnamefont
  {A.}~\bibnamefont {Planes}},\ }\href {\doibase 10.1103/PhysRevB.67.224406}
  {\bibfield  {journal} {\bibinfo  {journal} {Phys. Rev. B}\ }\textbf {\bibinfo
  {volume} {67}},\ \bibinfo {pages} {224406} (\bibinfo {year}
  {2003})}\BibitemShut {NoStop}%
\bibitem [{\citenamefont {Perkovi\ifmmode~\acute{c}\else \'{c}\fi{}}\ \emph
  {et~al.}(1999)\citenamefont {Perkovi\ifmmode~\acute{c}\else \'{c}\fi{}},
  \citenamefont {Dahmen},\ and\ \citenamefont {Sethna}}]{Perkovic1999}%
  \BibitemOpen
  \bibfield  {author} {\bibinfo {author} {\bibfnamefont {O.}~\bibnamefont
  {Perkovi\ifmmode~\acute{c}\else \'{c}\fi{}}}, \bibinfo {author}
  {\bibfnamefont {K.~A.}\ \bibnamefont {Dahmen}}, \ and\ \bibinfo {author}
  {\bibfnamefont {J.~P.}\ \bibnamefont {Sethna}},\ }\href {\doibase
  10.1103/PhysRevB.59.6106} {\bibfield  {journal} {\bibinfo  {journal} {Phys.
  Rev. B}\ }\textbf {\bibinfo {volume} {59}},\ \bibinfo {pages} {6106}
  (\bibinfo {year} {1999})}\BibitemShut {NoStop}%
\bibitem [{\citenamefont {Detcheverry}\ \emph {et~al.}(2003)\citenamefont
  {Detcheverry}, \citenamefont {Kierlik}, \citenamefont {Rosinberg},\ and\
  \citenamefont {Tarjus}}]{Detcheverry2003}%
  \BibitemOpen
  \bibfield  {author} {\bibinfo {author} {\bibfnamefont {F.}~\bibnamefont
  {Detcheverry}}, \bibinfo {author} {\bibfnamefont {E.}~\bibnamefont
  {Kierlik}}, \bibinfo {author} {\bibfnamefont {M.~L.}\ \bibnamefont
  {Rosinberg}}, \ and\ \bibinfo {author} {\bibfnamefont {G.}~\bibnamefont
  {Tarjus}},\ }\href {\doibase 10.1103/PhysRevE.68.061504} {\bibfield
  {journal} {\bibinfo  {journal} {Phys. Rev. E}\ }\textbf {\bibinfo {volume}
  {68}},\ \bibinfo {pages} {061504} (\bibinfo {year} {2003})}\BibitemShut
  {NoStop}%
\bibitem [{\citenamefont {Detcheverry}\ \emph {et~al.}(2004)\citenamefont
  {Detcheverry}, \citenamefont {Kierlik}, \citenamefont {Rosinberg},\ and\
  \citenamefont {Tarjus}}]{Detcheverry2004a}%
  \BibitemOpen
  \bibfield  {author} {\bibinfo {author} {\bibfnamefont {F.}~\bibnamefont
  {Detcheverry}}, \bibinfo {author} {\bibfnamefont {E.}~\bibnamefont
  {Kierlik}}, \bibinfo {author} {\bibfnamefont {M.~L.}\ \bibnamefont
  {Rosinberg}}, \ and\ \bibinfo {author} {\bibfnamefont {G.}~\bibnamefont
  {Tarjus}},\ }\href {\doibase 10.1021/la0488506} {\bibfield  {journal}
  {\bibinfo  {journal} {Langmuir}\ }\textbf {\bibinfo {volume} {20}},\ \bibinfo
  {pages} {8006} (\bibinfo {year} {2004})}\BibitemShut {NoStop}%
\bibitem [{\citenamefont {Phalippou}\ \emph {et~al.}(2005)\citenamefont
  {Phalippou}, \citenamefont {Woignier}, \citenamefont {Despetis},\ and\
  \citenamefont {Etienne-Calas}}]{Phalippou2005}%
  \BibitemOpen
  \bibfield  {author} {\bibinfo {author} {\bibfnamefont {J.}~\bibnamefont
  {Phalippou}}, \bibinfo {author} {\bibfnamefont {T.}~\bibnamefont {Woignier}},
  \bibinfo {author} {\bibfnamefont {F.}~\bibnamefont {Despetis}}, \ and\
  \bibinfo {author} {\bibfnamefont {S.}~\bibnamefont {Etienne-Calas}},\ }in\
  \href {http://hal.archives-ouvertes.fr/hal-00389867} {\emph {\bibinfo
  {booktitle} {{Handbook of sol-gel Science and Technology - Processing
  characterization and application}}}},\ \bibinfo {editor} {edited by\ \bibinfo
  {editor} {\bibfnamefont {H.}~\bibnamefont {Kozuka}}}\ (\bibinfo  {publisher}
  {Kluwer},\ \bibinfo {year} {2005})\ pp.\ \bibinfo {pages}
  {599--617}\BibitemShut {NoStop}%
\bibitem [{Note1()}]{Note1}%
  \BibitemOpen
  \bibinfo {note} {Aerogels grown in basic pH are an experimental realization
  of the numerical gels studied in ref.~\cite {Detcheverry2003,
  Detcheverry2004a}}\BibitemShut {NoStop}%
\bibitem [{\citenamefont {Melich}\ \emph {et~al.}(2011)\citenamefont {Melich},
  \citenamefont {Boltnev}, \citenamefont {Bonnet}, \citenamefont {Guyon},\ and\
  \citenamefont {Wolf}}]{Melich2011}%
  \BibitemOpen
  \bibfield  {author} {\bibinfo {author} {\bibfnamefont {M.}~\bibnamefont
  {Melich}}, \bibinfo {author} {\bibfnamefont {R.}~\bibnamefont {Boltnev}},
  \bibinfo {author} {\bibfnamefont {F.}~\bibnamefont {Bonnet}}, \bibinfo
  {author} {\bibfnamefont {L.}~\bibnamefont {Guyon}}, \ and\ \bibinfo {author}
  {\bibfnamefont {P.~E.}\ \bibnamefont {Wolf}},\ }\href {\doibase
  10.1007/s10909-010-0263-7} {\bibfield  {journal} {\bibinfo  {journal} {J. Low
  Temp. Phys.}\ }\textbf {\bibinfo {volume} {162}},\ \bibinfo {pages} {740}
  (\bibinfo {year} {2011})}\BibitemShut {NoStop}%
\bibitem [{\citenamefont {Cross}\ \emph {et~al.}(2007)\citenamefont {Cross},
  \citenamefont {Puech},\ and\ \citenamefont {Wolf}}]{Cross2007}%
  \BibitemOpen
  \bibfield  {author} {\bibinfo {author} {\bibfnamefont {B.}~\bibnamefont
  {Cross}}, \bibinfo {author} {\bibfnamefont {L.}~\bibnamefont {Puech}}, \ and\
  \bibinfo {author} {\bibfnamefont {P.~E.}\ \bibnamefont {Wolf}},\ }\href
  {\doibase 10.1007/s10909-007-9482-y} {\bibfield  {journal} {\bibinfo
  {journal} {J. Low Temp. Phys.}\ }\textbf {\bibinfo {volume} {{148}}},\
  \bibinfo {pages} {903} (\bibinfo {year} {{2007}})}\BibitemShut {NoStop}%
\bibitem [{Note2()}]{Note2}%
  \BibitemOpen
  \bibinfo {note} {Note that the pictures only evolve when $\Phi $ increases
  and not if condensation is paused. This is similar to the spin flips of the
  athermal RFIM which only take place when the field is swept.}\BibitemShut
  {Stop}%
\bibitem [{Note3()}]{Note3}%
  \BibitemOpen
  \bibinfo {note} {The choice of 135$^\circ $ rather than 45$^\circ $ minimizes
  the attenuation of the scattered light due to propagation through the bright
  region.}\BibitemShut {Stop}%
\bibitem [{Note4()}]{Note4}%
  \BibitemOpen
  \bibinfo {note} {More precisely, the signal drops to a slightly larger value
  than the filled state value. This is a second order scattering effect, due to
  the illumination of the filled aerogel by the recessing bright
  region.}\BibitemShut {Stop}%
\bibitem [{\citenamefont {P\'erez-Reche}\ \emph {et~al.}(2008)\citenamefont
  {P\'erez-Reche}, \citenamefont {Rosinberg},\ and\ \citenamefont
  {Tarjus}}]{Perez-Reche2008}%
  \BibitemOpen
  \bibfield  {author} {\bibinfo {author} {\bibfnamefont {F.~J.}\ \bibnamefont
  {P\'erez-Reche}}, \bibinfo {author} {\bibfnamefont {M.~L.}\ \bibnamefont
  {Rosinberg}}, \ and\ \bibinfo {author} {\bibfnamefont {G.}~\bibnamefont
  {Tarjus}},\ }\href {\doibase 10.1103/PhysRevB.77.064422} {\bibfield
  {journal} {\bibinfo  {journal} {Phys. Rev. B}\ }\textbf {\bibinfo {volume}
  {77}},\ \bibinfo {pages} {064422} (\bibinfo {year} {2008})}\BibitemShut
  {NoStop}%
\bibitem [{Note5()}]{Note5}%
  \BibitemOpen
  \bibinfo {note} {This quantity is measured, without any error due to
  attenuation, at the last pressure where the bright domain intersects the
  illuminated section.}\BibitemShut {Stop}%
\bibitem [{\citenamefont {Aubry}(2013)}]{Aubry2013b}%
  \BibitemOpen
  \bibfield  {author} {\bibinfo {author} {\bibfnamefont {G.~J.}\ \bibnamefont
  {Aubry}},\ }\emph {\bibinfo {title} {La condensation de l'h{\'e}lium 4 dans
  les a{\'e}rogels~: une r{\'e}alisation exp{\'e}rimentale du RFIM
  athermique}},\ \href@noop {} {\bibinfo {type} {Phd thesis}},\ \bibinfo
  {school} {Universit{\'e} de Grenoble} (\bibinfo {year} {2013})\BibitemShut
  {NoStop}%
\bibitem [{Note6()}]{Note6}%
  \BibitemOpen
  \bibinfo {note} {See Supplemental Material at [URL] for a description of the
  results on this light aerogel.}\BibitemShut {Stop}%
\bibitem [{\citenamefont {{Detcheverry}}\ \emph {et~al.}(2005)\citenamefont
  {{Detcheverry}}, \citenamefont {{Kierlik}}, \citenamefont {{Rosinberg}},\
  and\ \citenamefont {{Tarjus}}}]{Detcheverry2005a}%
  \BibitemOpen
  \bibfield  {author} {\bibinfo {author} {\bibfnamefont {F.}~\bibnamefont
  {{Detcheverry}}}, \bibinfo {author} {\bibfnamefont {E.}~\bibnamefont
  {{Kierlik}}}, \bibinfo {author} {\bibfnamefont {M.~L.}\ \bibnamefont
  {{Rosinberg}}}, \ and\ \bibinfo {author} {\bibfnamefont {G.}~\bibnamefont
  {{Tarjus}}},\ }\href {\doibase 10.1103/PhysRevE.72.051506} {\bibfield
  {journal} {\bibinfo  {journal} {Phys. Rev. E}\ }\textbf {\bibinfo {volume}
  {72}},\ \bibinfo {eid} {051506} (\bibinfo {year} {2005})},\ \Eprint
  {http://arxiv.org/abs/arXiv:cond-mat/0508267} {arXiv:cond-mat/0508267}
  \BibitemShut {NoStop}%
\end{thebibliography}%
%

  \end{document}